\documentclass{iaus}
\usepackage{graphicx}

\title[Spitzer Observations of TDGs] 
{Spitzer Observations of Tidal Dwarf Galaxies}

\author[Higdon \& Higdon]   
{Sarah J. U. Higdon$^{1,2}$ \& James L. Higdon$^1$%
}

\affiliation{$^1$ Physics Department, Georgia Southern University, Statesboro, GA 30460, 
\break $^2$ Astronomy Department, Cornell University, Ithaca, NY 14853
\break email: shigdon@georgiasouthern.edu}
\pubyear{2008}
\volume{244}  
\pagerange{xxx-xxx}
\date{?? and in revised form ??}
\jname{Dark Galaxies \& Lost Baryons}
\editors{J. I. Davies \& M. J. Disney, eds.}

\def\arcmin{$^{\prime}$}

\begin{document}

\maketitle

\begin{abstract}
We present Spitzer observations of Tidal Dwarf Galaxies (TDGs) in
three interacting systems: NGC~5291, Arp~105 \& Stephan's Quintet.
The spectra show bright emission from polyaromatic hydrocarbons
(PAHs), nebular lines and warm molecular hydrogen, characteristic of
recent episodes of star formation. The PAH emission that falls in the
IRAC 8.0 $\mu$m band leads to the TDGs having an extremely red IRAC color,
with [4.5] $-$ [8.0] $>$ 3. The emission from PAHs is characterized by
a model with mainly neutral 100-C PAH atoms.

\end{abstract}

\firstsection 
\section{Introduction}
Mergers of gas rich dusty spirals can form $>$100 kpc tidal tails
(Toomre \& Toomre 1972). Tidal Dwarf Galaxies (TDGs) condense out of
these tails, and may make up $1/3$ of the dwarf population in compact
groups (Hunsberger et al. 1996). Here we investigate the emission from
Polycyclic Aromatic Hydrocarbons (PAHs) in three TDG systems:\\ {\bf
  NGC 5291/Seashell} (western edge of Abell 3574). Disturbed
lenticular. Large fragmented HI ring in a mainly 8\arcmin eastern
arc. Two TDGs NGC 5291N \& NGC 5291S \& 13 TDG candidates (Spitzer
observations: Higdon, Higdon \& Marshall 2006).\\ {\bf Arp105/The
  Guitar} (Abell 1185). NGC 3561A(spiral) \& NGC
3561B(lenticular). Ambartzumian knot (TDG at tip of southern tail) \&
5 TDG candidates.\\ {\bf Stephan’s Quintet} (Arp 319) Prototype
compact group (Hickson 92).  NGC 7317(E), NGC 7318A(E), NGC 7318B (Sbc
pec), NGC 7319 (Sbc pec Sey 2) and NGC 7320 (Sd foreground
galaxy). Two 100 kpc parallel tails from NGC 7319 towards NGC
7320c. 40 kpc shock front in intra-group medium.  TDGs: SQ-A, SQ-B (Arp
1973), 13 TDG candidates (Hunsberger et al. 1996).

\section{TDGs Are PAH Rich \& Have Extreme IRAC Colors}
Spitzer Infrared Array Camera (IRAC) false color images showing the
PAH-rich TDGs are available at our web site\footnotemark [1] 
\footnotetext[1]{ http://cheller.phy.georgiasouthern.edu/$\sim$shigdon/tdg/tdg.html }.
The strong PAH emission is used to identify the TDGs in the IRAC
two-color diagram shown in Figure 1. The horizontal and vertical
lines mark the $[4.5]–-[8.0]$ and $[3.6]-–[4.5]$ color zero
points. Galaxies are sorted into 3 quadrants: Elliptical/Spheroidal
(bottom left). AGN/rising SED (top right). Star formation (bottom
right).  Compared to spiral galaxies and Blue Compact Galaxies (BCDs)
the TDGs occupy the extreme right with [4.5] $-$ [8.0] $>$ 3 (Higdon,
Higdon \& Marshall 2006). The knot in the ring galaxy, AM0644-74 is
consistent with the TDG color (Higdon \& Higdon 2007).

\begin{figure}
\centerline{
\scalebox{0.8} { 
\includegraphics[angle=90, height=4cm]{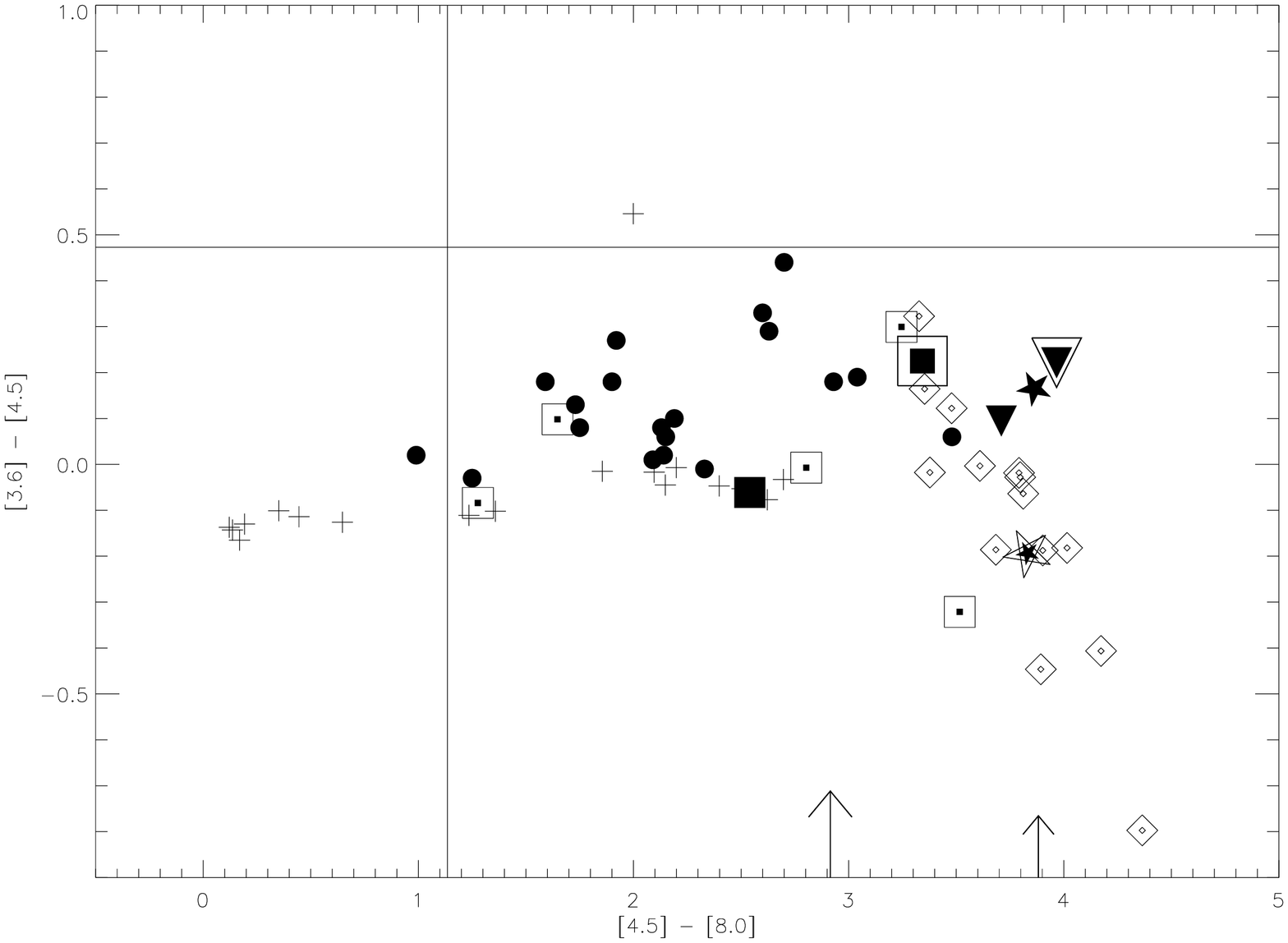}%
}}
\footnotesize
\caption{IRAC Two-Color Diagram: NGC 5291: NGC 5291N (solid star), NGC
  5291S (framed star), TDG candidates (framed diamonds). SQ-A \& SQ-B
  (triangles). Arp 105: TDG candidates (framed squares) \&
  Ambartzumian knot (framed large square). Star forming regions in the
  Cartwheel (large arrow) and in AM0644-74 (small arrow). For
  comparison: Hubble morphological sequence ( +, Pahre et al 2004);
  KISS BCD Sample (small filled circle, Rosenburg et al. 2005).}
\normalsize
\end{figure}

\section{TDGs Have  Mainly Neutral $\sim$100-C atom PAHs \& Warm H$_2$}

Spectra, which were obtained with the Spitzer low resolution Infrared
Spectrograph and extracted using SMART (Higdon et al. 2004), are rich
in PAHs, with H$_2$ in NGC 5291N \& S and NGC 7319A; and [SIV] in NGC
5291N \& Ambartzumian Knot.  The relative intensities of the
11.3/7.7$\mu$m and 6.2/7.7 $\mu$m PAH features were measured using PAHFIT
(Smith \& Draine 2007) and compared to models in Figure 16 from Draine
\& Li (2001).  The PAHs in NGC 5291 N\&S are consistent with $\sim$100
C-atoms. They are similar to the average PAH size of 27 normal
galaxies (Helou 2000), but with a slightly higher ion fraction. PAH
clumps are $\sim$4x bigger in Ambartzumian knot, whilst those in NGC
7319A are smaller and mostly neutral.

\section{Conclusions}

TDGs and knots of star formation in tidal streams tend to have
disturbed morphologies, with recent episodes of star formation, as
evidenced by emission from PAHs (modeled as mainly neutral 100-C
PAHs), nebular lines, and warm H$_2$. They can be identified by their
[4.5] $-$ [8.0] $>$ 3 color. As members of the Arecibo Legacy Fast
ALFA (ALFALFA) Undergraduate team, we will expand this study to
include new tidal features discovered in the ALFALFA Survey. Any found
without visible perturbers are potential candidates for small dark
matter halo interactions.

\small
\begin{acknowledgments}
Sarah thanks both the I.A.U. and the Physics Dept., Georgia Southern
University for travel grants.  The Spitzer Space Telescope is
operated by the Jet Propulsion Laboratory, Caltech, under NASA
contract 1407. Support for this work was provided by NASA, through
contract 1257184, issued by JPL/Caltech. We used the most excellent
NED.
\end{acknowledgments}

~


{\bf \noindent References}

 \noindent Draine, B. T. \& Li, A. 2001, ApJ, 551, 807 \\
Helou, G., Lu, N. Y., Werner, M. W., Malhotra, S., \& Silbermann, N. 2000 ApJ, 532, L21\\
Higdon, S. J. U.  et al. 2004 PASP, 116, 975\\
Higdon,  S. J. U., Higdon, J. L., \& Marshall, J. 2006, ApJ, 640, 768\\
Higdon,  S. J. U., Higdon, J. L. 2007 in prep. \\
Hunsberger, S. D., Charlton, J., \& Zaritsky, D. 1996 ApJ, 462, 50\\
Pahre, M. A., Ashby, M. L., Fazio, G. G., \& Willner, S. P. 2004 ApJS, 154, 235\\
Smith, J. D. T., \& Draine, B. T. et al., 2007, ApJ,656, 770\\
Toomre, A. \& Toomre, J. 1972, ApJ, 178, 623\\
   { } 

\end{document}